\documentstyle[aps]{revtex}

\begin{document}
\twocolumn
\title{Regular quantum interiors for black holes}
\author{Emilio Elizalde\cite{email,http}  and Sergi R. Hildebrandt}

\address{Instituto de Ciencias del Espacio (CSIC) \\ \&
Institut d'Estudis Espacials de Catalunya (IEEC/CSIC) \\
Edifici Nexus, Gran Capit\`a 2-4, 08034 Barcelona, Spain}

\maketitle
\begin{abstract}
A family of spacetimes  suitable for describing the
interior of a non-rotating black hole is constructed. The stress-energy
tensor is that of a spherically symmetric vacuum,
as commonly assumed nowadays.
The problem of matching the exterior with the interior region is
solved exactly, without using any massive shell nor having to
 restrict oneself to only asymptotically
well-behaved solutions, whatsoever. The main physical and
geometrical properties of the resulting black hole solutions are
described. As models for the interior the general solution found
includes, in particular,  two known previous attempts at solving
the problem.  Finally, effective macroscopic properties of the solution
are linked with quantization issues of the corresponding spacetime.
\end{abstract}

% DEFINITIONS

\def\lh{{\boldmath \hbox{$ \ell $}} }
\def\th{{\boldmath \hbox{$ \Theta $}} } % cobasis, Theta
\def\l{\Lambda}
\def\le{\Lambda_1} % Lambda, exterior %
\def\li{\Lambda_2} % Lambda, interior $
\def\tr{{\tilde r} }

\def\bce{\begin{center}}
\def\ece{\end{center}}
\def\beq{\begin{eqnarray}}
\def\eeq{\end{eqnarray}}
\def\ben{\begin{enumerate}}
\def\een{\end{enumerate}}
\def\ul{\underline}
\def\ni{\noindent}
\def\nn{\nonumber}
\def\bs{\bigskip}
\def\ms{\medskip}

Work on collapsing bodies has shown \cite{funda,chandra} that there
exists a                      mass   above  which the collapse of an
object and the appearance of a singularity are unavoidable. In
searching for new solutions, with the aim to circumvent
 such conclusion, it has been
claimed that other quantum effects, specially those associated with
the quantum vacuum, might actually yield regular solutions everywhere
inside the object. Indeed,
to stop gravitational collapse of a very massive body,
the final solution should develop a static region that
coincides with the core of the object.
The strong energy conditions \cite{he} must be
violated inside the final body, in order to make the singularity
avoidable. Even tough any realistic object may differ significantly
from spherical symmetry at the beginning of the collapse, there are
several mechanisms by which, close to the final state, spherical
symmetry holds, both for the exterior as well as for the interior regions
\cite{ip,letters}, to the point where the classical notions
of space and time loose their meaning.

The general expression for the matter-energy content of such
contributions, for spherically symmetric bodies, has the form
$ \rho + p = 0 $, and $ p_2 = p_3 $ \cite{zn,dymni}, where
$ \rho $ is the mass-energy density,
$ p $ the radial stress (or pressure) and $ p_2 = p_3 $ the
transversal stresses (pressures) measured by any local observer
(proper reference frame).

We will here deal with the family of maximal spherically symmetric
spacetimes expanded from flat spacetime by a geodesic radial null
one-form (GRNSS spaces, see \cite{magli,rqibh,pakss}), a subfamily 
of Kerr-Schild (KS) metrics. They are given by
\beq ds^2 = ds^2_{\eta} + 2 H(r) \lh \otimes \lh, \eeq
where $ ds^2_{\eta} $ stands for the flat spacetime metric, H is an
arbitrary
function of $ r $, the radial coordinate of the spherical symmetry
---defined in
some open region of the manifold--- and $ \lh $ is a geodesic radial
null
one-form. These metrics
include {\em all} spherically symmetric spacetimes
which have a static region and satisfy the previous energy-matter
constraints. Another expression
for the family is (with signature ($-+++$))
\beq
ds^2 &=& -(1-H) dt^2 + 2 H dt\,dr + (1+H) dr^2 \nn \\ && + r^2 \bigl(
d\theta^2 +{\sin^2}\theta \, d\varphi^2 \bigr),
\eeq
 with
$ \lh = (1/\sqrt{2})(dt + dr) $. The other possibility, i.e. $
\lh = (1/\sqrt{2})(dt - dr) $ yields the same physical results. Since $
H=H(r) $,
$ \partial_t $ is an integrable Killing vector. 

%%Particularly,  for $ H<1
%%$ it
%%is time-like, for $ H = 1 $ null, and for $ H > 1 $ space-like. 
The above choice avoids coordinate problems near their possible
horizons, e.g. when $ H = 1 $.

In the region where $ H < 1 $, the existence of such integrable
Killing vector allows to write the whole family of metrics in an explicitly
static form,
\beq
 ds^2 = - (1-H){dt_s}^2 + {1 \over  1- H} dr^2 + r^2 \bigl(
d\theta^2 + {\sin^2}\theta \, d\varphi^2 \bigr),\nn
\eeq
where $ dt_s $ is related to $ dt $ by
$ dt_s = dt -  [H/(1- H)] dr. $
This last expression for the family looks as a generalization of the
well known
Schwarzschild metric and may help to identify the class of spacetimes
we are dealing with, \cite{ms}.

In order to recover the natural scheme of collapsed objects, one must
consider the matching of two spherically symmetric solutions of the
Einstein equations. Here, the interior solution will belong to the class
GRNSS
while the exterior one may be {\em any} solution for a non-rotating
classical black hole. One can consider e.g. a charged, Reissner-Nordstr\"{o}m
solution, a
Schwarzschild solution \cite{chandra,mtw,wald}, or
even a cosmological background including a cosmological constant,
\cite{kramer}. Regardless of the choice of
exterior metric, we have the additional advantatge
that {\em all} these exteriors solutions also belong
to the GRNSS family of
spaces. As a consequence, the matching becomes simple.

%%Since we aim at describing a global solution, representing a central
%%object matched with its surroundings, which will hold ``for all times'' (here
%%``time'' is referred to the one measured by any observer in the static
%%exterior region), the matching hypersurface must be spatia everywhere. 
There are in fact many possible choices for the
equations defining the matching hypersurface \cite{rqibh},
the most interesting  is simply to choose the hypersurface of
constant $ r $. Other possibilities (avoiding the singularity) should
converge to this choice.

The corresponding matching conditions translate into: 
%% \cite{rqibh}

$ [H]=0, [H'] =0, $
where $ [f] = f_{\rm ext} - f_{\rm int} $ and the prime stands for the
radial ordinary derivative. Their physical interpretation is direct, the first is the continuity of the mass function, while the second assures the continuity of radial pressures.

%A consequence of the matching conditions is the impossibility of directly
%joining Schwarzschild and de Sitter solutions. For the
%Schwarzaschild solution one has $ H = 2m/r $, where $ m $ is the
%gravitational mass of the object; for the de Sitter solution, $ H = \l
%r^2/3 $, where $ \l $ is a constant, e.g. the cosmological
%constant. From the matching conditions one readily sees
%that the system is incompatible. Nevertheless, the
%possibility remains of doing the match at a null hypersurface, as a limiting
%case. A detailed study of this issue for two GRNSS
%spaces shows that this is eventually impossible \cite{rqibh}, see also
%\cite{ip}, contrary to previous claims, \cite{sz}.

%Having obtained the necessary and suffcient conditions that allow to join
%two arbitrary metrics of the GRNSS family, the next
%task is  to analyze if these natural scheme can yield the expected
%description of a regular interior for a non-rotating black hole. It is
%worth emphasizing that we could also handle singular spacetimes,
%\cite{ip,letters}, just by changing the matching hypersurface.
%But what we consider here is the possibility to {\it avoid} the central
%singularity.
To
%this end
avoid the central  singularity
we could impose a specific form of the
matter-energy content under some general accepted scheme of
quantum behaviour close to the origin. However it is
more interesting to demand only that the solution be regular at the
origin and study the main (universal) properties of all such candidates. In this
way we shall extract some general information, unbiassed by our
prejudices on the ``ultimate state of matter''.

A first remarkable consequence is  that, regardless of the
chosen specific model for the interior, {\em all}  become
isotropic near the origin, i.e. $ p_2  = p $ at the origin. This
property has been always imposed ever since the incorporation of high
energy quantum effects in the field, and is tacitly assumed in all
previous attempts. Here it {\it directly 
follows} from the GRNSS defining conditions,
for which we only demanded $ \rho = -p $, $ p_2 = p_3 $,
and partial staticity, but not necessarily isotropy.
In consequence, we do get a {\em smooth} transition
from the exterior phase to the central one, as expected by previous
authors, see e.g. the conclusions in \cite{bp}. Moreover -and this is also a
very
important point-  the need for a (thin) layer has now disappeared, \cite{magli}. In
summary, the matching of two  GRNSS spaces recovers, in an absolutely
natural way, {\em
all} the features of existing collapse schemes for massive objects and
extends this issue to the domain of new highly energetic quantum processes.

For a solution of the GRNSS class to be everywhere
regular, it suffices that $ H_2(r=0) = 0 $, and $ H'_2(r=0) = 0 $,
where  label 2 refers to the interior solution. This result on the
regularity of the inner solution is based on the analysis of the
Riemannian invariants of the metrics. For instance, $ {\cal R}^2
\equiv R_{\mu \nu \lambda \rho} R^{\mu \nu \lambda \rho} $ yields
$  (H'')^2 + 4 (H'/r)^2 + 4 (H/r^2)^2 $. Any regular interior model
develops here
 a maximum limiting value for all these invariants, as also expected by
several authors, e.g. \cite{markov}.

Another important consequence of the general regular models is that
the ``mass function'', defined as usual by $ m(r) =r(1-g_{\rm
static}^{rr})/2 $, is  $ rH(r)/2 $ and becomes smaller and smaller as
the origin is approached, until it vanishes. Thus {\em all} these
models also yield the result that they are asymptotically free in
the sense that the gravitational charge $ m(r) $ vanishes at the origin.
In \cite{fmm}, Sect. II,  results were claimed to be possibly
valid also for any asymptotically free model of the interior core.
Here we prove rigorously that {\em all} regular-at-the-origin GRNSS spaces
are asymptotically free and share those desirable physical properties.

%Despite the fact that all these results are completely independent
%of the matter-energy content around the origin, a most important
%issue in finding specific solutions
%among all the possibilities has to do with the
%physical behaviour near the origin. Isotropization has been proven to
%be generic within any model. However this isotropization is till now
%of a geometrical character. We must specify what type
%of physical isotropization is expected.

A number of papers have dealt with this issue, looking for
plausible energy contributions near the origin.
Thus, in \cite{glinner} it is advocated for a de Sitter
core, which has been hitherto the most studied situation, $
H = \l r^2/3 $, clearly satisfying the regularity conditions. But there is
still a broad open window for other alternatives. As yet there is
no quantum gravity theory available and therefore an assumption
regarding the precise form of the stress-energy tensor at the origin cannot
be made. However there are impelling arguments in favor
of such a behaviour near the origin and the de Sitter choice will then
correspond to the limiting case.

For the exterior region, one usually assumes a Schwarz\-schild black
hole solution, i.e. a non-charged one (the case of a
%%an electrically or magnetically
charged black hole can be easily recovered from our
results below). Observational evidence \cite{astro},
as well as theoretical arguments \cite{mtw}, lead, on the other hand,
to the conclusion that
non-charged black holes are the most common ones in our universe.
Nevertheless, one can also consider the effect of adding a
cosmological background to the Schwarzschild solution in terms of a
cosmological term, and/or also other contributions, as for
instance  quantum contributions coming from the vacuum polarization of
the exterior region close to the matching hypersurface. Generalically
one will have a relation of the type $ H_2 (r\sim 0) \sim \li r^2/3 $,
and $ H_1=H_1(r, m_1,\le, \{\alpha^1_i\}) $, where $ m_1 $ is the
gravitational mass as measured by an external observer, $ \le $  the
(possible) cosmological term of the outer region, and $ \{ \alpha^1_i\}
$  a set of parameters describing the strength of other effects, such
as the charge of the black hole, vacuum polarization of the exterior
solution, etc. By virtue of the matching conditions we will have
schematically $ H_2 =H_2(r; m_1, \le, \{\alpha^{1,2}_i\},R) $, where $ R
$ is the value at which the two solutions match.  The energy-matter
density for any metric of the GRNSS spaces reads (G=c=1)$
8 \pi \rho_2 = (H_2r)'/r^2 $. Thus $ \rho_2 $ will have an analogous
dependence which, once imposed the criteria
$ 8 \pi \rho_2 (r=0) = \li $, will yield a relationship of the type $ \li =
F(R,m_1,\le,\{\alpha^{1,2}_i\}) $. In order to determine $ R $ in terms of
the rest of the parameters, one has to consider a specific model for
the interior.
%This is by no means a handicap, but it is a desired result.
As in \cite{funda,chandra,mtw},
%%Comparing with the related problem of spherical stars
we encounter
%% the result is analogous. In both cases
a big set of allowed models for the
interiors of the objects given a common exterior and the ultimate task
is that of finding  realistic physical models for these interiors.
%So far no realistic model is known for the quantum behaviour of the interior
%of a spherical body
%(due in part to the poor knowledge we have of the interface
%between classical gravity and high-energy quantum effects).
%Actually, two specific models for the matter-energy profile inside
%such body have been proposed, but no quantum field justification for them has
%been found.
In \cite{rqibh} we  perform in detail the calculations of the
corresponding models in our formulation, as well as of a whole set of
new proposals,
in order to investigate the universality of the conclusions
derived from these particular examples.
%The result of this analysis has shown a number of surprinsing similarities
%in all cases,
%even though the collection of interiors and exteriors is very widespread.

The first known model\cite{ip,fmm} has its analogue here taking $ H_1=
2m_1/r-(1/3)\times\allowbreak
(\alpha m_1/r^2)^2 $, where $ \alpha^2 $ is a number related with the
number and types of the quantized fields, being of order unity in
Planckian units, and $ H_2 = r^2/3B(B+Cr^3) $, where $ B =
\alpha/(6-\alpha^2 m_1/R^3) $ and $ C = (2/\alpha m_1)[1-3/(6-
\alpha^2 m_1/R^3) ] $.  $ H_1 $ takes into account corrections
coming from the polarization of the vacuum that is expected to be
proportional to $ {\cal R}^2 $.
The second known model\cite{bd,dymni} is retrieved by setting $ H_1 = 2m_1/r
+\le r^2/3 $, and $ H_2 = (R^2/3)\{(\beta/\alpha)[\exp{(-
\alpha\tr^3)}-1]/\tr -\gamma \tr^2\} $, where $ \gamma = -
\beta\exp{(-\alpha)} - \le $, and $ \alpha $, $ \beta $ are solutions of
trascendental equations. A valid approximate expression is $
\alpha \simeq (\le-\li)R^3/6m_1 $, and 
$ \beta \simeq \le-\li $.
The other set of new models considered is described by $ H_1 =
2m_1/r +\le r^2/3 $, and $ H_2 = b_2 r^2 + b_N r^N $, with $ N \ge
3 $, where $ b_2 $ is given by $ (2m_1/R)(N+1)/(N-2) + \le R^2/3 $,
and $ b_N= -6m_1/R(N-2) $.

Obviously the value of $ R $ will depend in general on the
regularization scale at which the usual concepts of space and
time lose their sense. To fix ideas, let us assume that the
regularization scale is of Planckian order (other choices are to be
found in \cite{rqibh}). The result is that in {\em all} the preceding
models $ R $ remains constrained to the range $ 2k \root{3}\of{M} 
(10^{-20} {\rm cm}) $, with $ k \sim [1,10] $, and $ M $ defined by $
m_1/m_{\odot} $, being $ m_{\odot} $ the solar mass.

There are some issues       worth  mentioning about
the models. First, all possess a matching $ R $ wich is
well above the regularization scale for any astrophysical
object ($ m \sim m_{\odot} $). For instance, in the case of the
Planck scale, the result is  $ R/L_{\rm Pl} \sim 10^{12} >>1 $.
This boosts our confidence on the plausibility of the  solutions found,
since they are indeed very far from the scales where the
applied scheme of continuous space an time loses its sense. On the
other hand, these objects exhibit in general two main horizons. The first is
the usual event horizon, that coincides with the typical one of a black
hole, located at $ r = 2m_1 $. The second is a (non-global)
Cauchy horizon (CH) \cite{he,chandra} near the origin, where $ H_2 = 1$. For $
r<r_{\rm Cauchy} $ the spacetime becomes static and we enter into
the de Sitter core. In addition to these ``local'' horizons,
we could also have others, coming from the cosmological background
\cite{dymni,ds,rqibh} (see also Fig. 1 in \cite{rqibh}). 
The total number of horizons affects the so-called stability of 
the inner CH. This question has received a lot of attention in the 
literature, where non-stationary perturbations of the static, or 
stationary, models show, at the classical or semiclassical level, that 
the CH may become unstable, e.g. \cite{letters}. Yet this (dynamical) 
instability is weak \cite{ori}, so that the formation of the core 
is still possible.
Moreover, the strong energy conditions
\cite{he,wald} are violated far enough from the regularization scale, so
that eventually the singularity is not created. This is in accordance
with known singularity theorems, which need the strong energy condition to
be
satisfied near the singularity. 
%The general structure of the  models considered
%is depicted in the Penrose (conformal) diagram of Fig. 1.
% Put the static and non-stationary regions, etc., without extending
% it.
% The extension can be led for the following section.

Also worth considering is the stability of the
solutions:
will they still exist in case of changes in
the parameters such as $ m_1 $, because of mass
accretion, the regularization scale, etc.? If our models were
highly dependent on some of these parameters, then the regular
solution would be valid only for a ``fine tunned configuration''.
Along the lines of \cite{bp} we have proven that our
models are stable.
%Thus for any smooth change in the parameters our
%results change also smoothly.
Another issue concerns the creation of a closed world inside the
collapsed object. This possibility can be interpreted as the creation of
a new (inflationary) macroscopic universe connected with the
collapsed body via its core (see Fig. 2 in \cite{rqibh}).
%The spacetime diagram for such a
%possibility is depicted in Fig. 2.
In the two models studied in
\cite{fmm,dymni,ds} this property holds. Here,
because of the properties of asymptotically free cores and the
existence of a non-global Cauchy horizon this possibility is always
present, see also \cite{borde}.

We note that all these models keep the matching hypersurface
inside the event horizon. In general, this has to be indeed
the case for any collapsed body exceding approximately three solar
masses. Of course, other objects with a lower mass can also form a
black hole provided its matter-energy content is sufficiently
compressed. Therefore, in order to analyze whether the
singularity of a black hole is avoidable, we must
descend to the physics inside the event horizon and study matter-energy
contents that will allow for the necessary violation of the energy
conditions to stop the creation of the singularity. So far the only
candidates are those coming from quantum field theory, because we
fairly know that matter and energy become quantized under the physical
conditions prevailing in these objects. Even though a  theory of
quantum
gravity is yet unavailable and, therefore, no exact solution is still
known for such merge, there are some results that point towards a
plausibility of the avoidance of singularities. The reasons for such
confidence are twofold. First, concerning
the type of energy contributions  to be expected near the origin
of the object, several studies \cite{glinner}  lead
to the conclusion that the quantum energy of the vaccum acts as a
negative stress and could remove the appearance of the singularity.
This crucial point is explicitly present in our geometrical description
in terms of the isotropization near the origin and through its
association with an internal $ \l $. These first results touch the
most intrincate problem, namely that of the plausible physics near the
regularization scale, but let us recall that different such possibilities
can be easily incorporated into our scheme, 
e.g. stringy black holes \cite{rqibh}.

Second, there is the region far away from the renormalization scale,
where one trusts the semiclassical theory of gravity \cite{bd} to be an
accurate enough description of the involved physics. In fact this can be
proven for any object satisfying $ m_1 >> M_{\rm Reg} $, $
R/L_{\rm Reg} >> 1 $.
%% will be much greater than one.
For instance, for an
astrophysical black hole and a regularization scale of the order of the
Planck scale, we obtain $ R/L_{Pl} \sim 10^{12} $, $ R \sim
10^{-20} {\rm cm} $, see also \cite{ip}. Thus, the 
fundamental question appears
of whether any of these models corresponds to a solution of the
semiclassical equations of gravity, at least in the  domain where $
r/ L_{\rm Reg} $ is big enough. As far, no one has found any 
such suitable
model,
however under very
general assumptions we have considered here {\em all} the possible
candidates
with spherical symmetry which have a static region and exhibit the
expected behaviour for the quantum stress-energy tensor  of a collapsed
body. Thus, if any particular solution is found in the future,
we believe that it will be inside this family.

To find the sources for the models here
presented is not easy, but this will be a  necessary step
to undertake, to show the plausibility 
of avoidance of the singularities.
To reach this goal  within our scheme, one 
needs in fact to quantize the sources.
The symmetries of the matter-energy tensor, i.e. $
\rho + p = 0 $, and the one coming from spherical symmetry, $
p_2 =p_3 $ are fundamental in order to check which type of
quantum fields could correspond to their sources. Taking into account the
spherical symmetry of the spacetime and, consequently, of the
Einstein tensor, one easily checks that all classical fields adapted to
the spherical symmetry satisfy this condition, see for instance Sect. 3.8
of
\cite{bd}. This bonus was indeed expected, because spherical
symmetry is a basic geometrical symmetry in the scheme. On the other hand,
the other eq., $ \rho + p = 0 $, can be
rewritten in a general covariant way, without referring to any special
set of observers. Its expression is then $ T_{ll} \equiv T_{\mu
\nu}l^{\mu} l^{\nu} = 0 $, where $ \vec \ell $ is the geodesic radial
null direction characteristic of the GRNSS spaces.
In the scalar case, for instance, one obtains:
$ T_{ll} = (1-2\xi){\phi'}^2 - 2 \xi\phi\phi'', $
where $ ()' \equiv {\vec \ell}() $, and $ \xi $ is a constant
representing the coupling between the scalar and the gravitational
field. If one imposes free scalar field equations, one finds no
solution. The end result is that no free classical field satisfies this
requirement. On the other hand, de Sitter spacetime has $ <T_{\alpha \beta}>
= -\l g_{\alpha \beta}$, thus $ <T_{ll}> = 0 $, what means that the
cancellation of $ T_{ll} $ is indeed due to quantum regularization of the
classical field. A starting program of quantization of GNRSS spaces has led
to the result that $ T_{ll} $ plays essentially the same role as $
T^{\lambda}_{\lambda} $ in related conformal spacetimes, and that the
vanishing of $ <T_{ll} >$ is most likely an anomaly effect,
also influenced by
the mass of the field and its coupling to the gravitational field. 
Other sources are to be found within effective 
actions of M-theory or string 
theory, see e.g. \cite{ayon}.

Another ---maybe less fundamental--- way of attaining the expected
quantum corrections
would be to consider some anisotropic version of  
spherical collapse
\cite{zn} and to perform the calculations associated with  
vacuum polarization
effects. The form of the GNRSS spaces allows in 
fact for a direct calculation of
such effects. This is a different physical point of 
view from the
one advocated in \cite{letters}, where the authors have been
mainly concerned with the radiative processes inside 
the black hole. In our
analysis, the results for the de Sitter  and  Schwarzschild cases yield
well-kown statements that have been duely taken into account in the
models presented above.

We conclude that it seems natural that a  suitable 
spacetime inside the GNRSS
family will correspond to a physically realistic 
quantum model of the interior of a BH,
free of singularities. On top of this conclusion, we recall again 
that other
possibilities, even different singular models, are also 
contained in the GNRSS
family
---as long as they satisfy the usual stress-energy conditions 
for a spherical
vacuum. In all, a remarkable amount of plausible 
realizations within this
family. An extension to rotating BH in a similar framework has been recently 
 considered\cite{burinskii}.
%\ms

Work supported by DGICYT (Spain), project
PB96-0925 and  CIRIT (Generalitat de Catalunya), contract 1999SGR-00257.

\end{document}